\documentclass[reprint,superscriptaddress,showpacs,secnumarabic,amssymb, nobibnotes, nofootinbib, aps, prl,floatfix]{revtex4-1}
\pdfoutput=1

\usepackage{graphicx}
\usepackage{epstopdf}
\usepackage{color}
\usepackage{amsmath}
\usepackage{hyperref}
\usepackage{bm}
\usepackage[dvipsnames]{xcolor}
\usepackage[left]{lineno}

\newcommand{\open}{{<\kern -0.3 em{\scriptscriptstyle )}}}

\newcounter{comment}

\begin{document}
%\linenumbers

\title{Beam spin asymmetry in semi-inclusive electroproduction of a hadron pair}

\newcommand*{\ANL}{Argonne National Laboratory, Argonne, Illinois 60439}
\newcommand*{\ANLindex}{1}
\affiliation{\ANL}
\newcommand*{\CANISIUS}{Canisius College, Buffalo, NY}
\newcommand*{\CANISIUSindex}{2}
\affiliation{\CANISIUS}
\newcommand*{\CMU}{Carnegie Mellon University, Pittsburgh, Pennsylvania 15213}
\newcommand*{\CMUindex}{3}
\affiliation{\CMU}
\newcommand*{\CUA}{Catholic University of America, Washington, D.C. 20064}
\newcommand*{\CUAindex}{4}
\affiliation{\CUA}

\newcommand*{\FERMI}{Centro Fermi - Museo Storico della Fisica e Centro Studi e Ricerche “Enrico Fermi", Rome, Italy}
\newcommand*{\FERMIindex}{4}
\affiliation{\FERMI}

\newcommand*{\SACLAY}{IRFU, CEA, Universit\'{e} Paris-Saclay, F-91191 Gif-sur-Yvette, France}
\newcommand*{\SACLAYindex}{5}
\affiliation{\SACLAY}
\newcommand*{\CNU}{Christopher Newport University, Newport News, Virginia 23606}
\newcommand*{\CNUindex}{6}
\affiliation{\CNU}
\newcommand*{\UCONN}{University of Connecticut, Storrs, Connecticut 06269}
\newcommand*{\UCONNindex}{7}
\affiliation{\UCONN}
\newcommand*{\DUKE}{Duke University, Durham, North Carolina 27708-0305}
\newcommand*{\DUKEindex}{8}
\affiliation{\DUKE}
\newcommand*{\DUQUESNE}{Duquesne University, 600 Forbes Avenue, Pittsburgh, PA 15282 }
\newcommand*{\DUQUESNEindex}{9}
\affiliation{\DUQUESNE}
\newcommand*{\FU}{Fairfield University, Fairfield CT 06824}
\newcommand*{\FUindex}{10}
\affiliation{\FU}
\newcommand*{\FERRARAU}{Universita' di Ferrara , 44121 Ferrara, Italy}
\newcommand*{\FERRARAUindex}{11}
\affiliation{\FERRARAU}
\newcommand*{\FIU}{Florida International University, Miami, Florida 33199}
\newcommand*{\FIUindex}{12}
\affiliation{\FIU}
\newcommand*{\FSU}{Florida State University, Tallahassee, Florida 32306}
\newcommand*{\FSUindex}{13}
\affiliation{\FSU}

\newcommand*{\GUAMEX}{Instituto de F\'isica, Universidad Nacional Aut\'onoma de M\'exico Apartado Postal 20-364, Ciudad de M\'exico 01000, M\'exico}
\newcommand*{\GUAMEXindex}{15}
\affiliation{\GUAMEX}

\newcommand*{\GWUI}{The George Washington University, Washington, DC 20052}
\newcommand*{\GWUIindex}{14}
\affiliation{\GWUI}
\newcommand*{\ISU}{Idaho State University, Pocatello, Idaho 83209}
\newcommand*{\ISUindex}{15}
\affiliation{\ISU}
\newcommand*{\INFNFE}{INFN, Sezione di Ferrara, 44100 Ferrara, Italy}
\newcommand*{\INFNFEindex}{16}
\affiliation{\INFNFE}
\newcommand*{\INFNFR}{INFN, Laboratori Nazionali di Frascati, 00044 Frascati, Italy}
\newcommand*{\INFNFRindex}{17}
\affiliation{\INFNFR}
\newcommand*{\INFNGE}{INFN, Sezione di Genova, 16146 Genova, Italy}
\newcommand*{\INFNGEindex}{18}
\affiliation{\INFNGE}
\newcommand*{\INFNRO}{INFN, Sezione di Roma Tor Vergata, 00133 Rome, Italy}
\newcommand*{\INFNROindex}{19}
\affiliation{\INFNRO}
\newcommand*{\INFNTUR}{INFN, Sezione di Torino, 10125 Torino, Italy}
\newcommand*{\INFNTURindex}{20}
\affiliation{\INFNTUR}
\newcommand*{\INFNPAV}{INFN, Sezione di Pavia, 27100 Pavia, Italy}
\newcommand*{\INFNPAVindex}{21}
\affiliation{\INFNPAV}
\newcommand*{\ORSAY}{Universit'{e} Paris-Saclay, CNRS/IN2P3, IJCLab, 91405 Orsay, France}
\newcommand*{\ORSAYindex}{22}
\affiliation{\ORSAY}
\newcommand*{\Juelich}{Institute fur Kernphysik (Juelich), Juelich, Germany}
\newcommand*{\Juelichindex}{23}
\affiliation{\Juelich}
\newcommand*{\JMU}{James Madison University, Harrisonburg, Virginia 22807}
\newcommand*{\JMUindex}{24}
\affiliation{\JMU}
\newcommand*{\KNU}{Kyungpook National University, Daegu 41566, Republic of Korea}
\newcommand*{\KNUindex}{25}
\affiliation{\KNU}
\newcommand*{\LAMAR}{Lamar University, 4400 MLK Blvd, PO Box 10009, Beaumont, Texas 77710}
\newcommand*{\LAMARindex}{26}
\affiliation{\LAMAR}
\newcommand*{\MIT}{Massachusetts Institute of Technology, Cambridge, Massachusetts  02139-4307}
\newcommand*{\MITindex}{27}
\affiliation{\MIT}
\newcommand*{\MISS}{Mississippi State University, Mississippi State, MS 39762-5167}
\newcommand*{\MISSindex}{28}
\affiliation{\MISS}
\newcommand*{\ITEP}{National Research Centre Kurchatov Institute - ITEP, Moscow, 117259, Russia}
\newcommand*{\ITEPindex}{29}
\affiliation{\ITEP}
\newcommand*{\UNH}{University of New Hampshire, Durham, New Hampshire 03824-3568}
\newcommand*{\UNHindex}{30}
\affiliation{\UNH}
\newcommand*{\NSU}{Norfolk State University, Norfolk, Virginia 23504}
\newcommand*{\NSUindex}{31}
\affiliation{\NSU}
\newcommand*{\OHIOU}{Ohio University, Athens, Ohio  45701}
\newcommand*{\OHIOUindex}{32}
\affiliation{\OHIOU}
\newcommand*{\ODU}{Old Dominion University, Norfolk, Virginia 23529}
\newcommand*{\ODUindex}{33}
\affiliation{\ODU}
\newcommand*{\RPI}{Rensselaer Polytechnic Institute, Troy, New York 12180-3590}
\newcommand*{\RPIindex}{34}
\affiliation{\RPI}
\newcommand*{\URICH}{University of Richmond, Richmond, Virginia 23173}
\newcommand*{\URICHindex}{35}
\affiliation{\URICH}
\newcommand*{\ROMAII}{Universita' di Roma Tor Vergata, 00133 Rome Italy}
\newcommand*{\ROMAIIindex}{36}
\affiliation{\ROMAII}
\newcommand*{\MSU}{Skobeltsyn Institute of Nuclear Physics, Lomonosov Moscow State University, 119234 Moscow, Russia}
\newcommand*{\MSUindex}{37}
\affiliation{\MSU}
\newcommand*{\SCAROLINA}{University of South Carolina, Columbia, South Carolina 29208}
\newcommand*{\SCAROLINAindex}{38}
\affiliation{\SCAROLINA}
\newcommand*{\TEMPLE}{Temple University,  Philadelphia, PA 19122 }
\newcommand*{\TEMPLEindex}{39}
\affiliation{\TEMPLE}
\newcommand*{\JLAB}{Thomas Jefferson National Accelerator Facility, Newport News, Virginia 23606}
\newcommand*{\JLABindex}{40}
\affiliation{\JLAB}
\newcommand*{\UTFSM}{Universidad T\'{e}cnica Federico Santa Mar\'{i}a, Casilla 110-V Valpara\'{i}so, Chile}
\newcommand*{\UTFSMindex}{41}
\affiliation{\UTFSM}
\newcommand*{\INSUBRIA}{Universit\`{a} degli Studi dell'Insubria, 22100 Como, Italy}
\newcommand*{\INSUBRIAindex}{42}
\affiliation{\INSUBRIA}
\newcommand*{\BRESCIA}{Universit\`{a} degli Studi di Brescia, 25123 Brescia, Italy}
\newcommand*{\BRESCIAindex}{43}
\affiliation{\BRESCIA}
\newcommand*{\GLASGOW}{University of Glasgow, Glasgow G12 8QQ, United Kingdom}
\newcommand*{\GLASGOWindex}{44}
\affiliation{\GLASGOW}
\newcommand*{\YORK}{University of York, York YO10 5DD, United Kingdom}
\newcommand*{\YORKindex}{45}
\affiliation{\YORK}
\newcommand*{\VT}{Virginia Tech, Blacksburg, Virginia   24061-0435}
\newcommand*{\VTindex}{46}
\affiliation{\VT}
\newcommand*{\VIRGINIA}{University of Virginia, Charlottesville, Virginia 22901}
\newcommand*{\VIRGINIAindex}{47}
\affiliation{\VIRGINIA}
\newcommand*{\WM}{College of William and Mary, Williamsburg, Virginia 23187-8795}
\newcommand*{\WMindex}{48}
\affiliation{\WM}
\newcommand*{\YEREVAN}{Yerevan Physics Institute, 375036 Yerevan, Armenia}
\newcommand*{\YEREVANindex}{49}
\affiliation{\YEREVAN}

\newcommand*{\NOWISU}{Idaho State University, Pocatello, Idaho 83209}
\newcommand*{\NOWBRESCIA}{Universit\`{a} degli Studi di Brescia, 25123 Brescia, Italy}
\newcommand*{\NOWJLAB}{Thomas Jefferson National Accelerator Facility, Newport News, Virginia 23606}
 %%%%%%%%%%%%%%% END OF Latex Macros for institute addresses  %%%%%%%%%%%%%%%%%%%%%%%%% 

 %%%%%%%%%%%%%%% END OF Latex Macros for institute addresses  %%%%%%%%%%%%%%%%%%%%%%%%% 

\author {M.~Mirazita} 
\affiliation{\INFNFR}
\author {H.~Avakian}
\affiliation{\JLAB}
\author{A.~Courtoy}
\affiliation{\GUAMEX}
\author{S.~Pisano}
\affiliation{\FERMI}

\author {S. Adhikari} 
\affiliation{\FIU}
\author {M.J.~Amaryan} 
\affiliation{\ODU}
\author {G.~Angelini} 
\affiliation{\GWUI}
\author {H.~Atac} 
\affiliation{\TEMPLE}
\author {N.A.~Baltzell} 
\affiliation{\JLAB}
\author {L. Barion} 
\affiliation{\INFNFE}
% ...

\collaboration{The CLAS Collaboration}
\noaffiliation
%%%%%%%%%%%%%%%%%%%%%%%%%%%%%%%%%%%%% END LIST OF AUTHORS %%%%%%%%%%%%%%%%%%%%%%%

\date{\today}

\begin{abstract}
\noindent A first measurement of the longitudinal beam spin asymmetry $A_{LU}$ in the semi-inclusive electroproduction of pairs of charged pions is reported. $A_{LU}$ is a higher-twist observable and offers the cleanest access to the nucleon twist-3 parton distribution function $e(x)$.
Data have been collected in the Hall-B at Jefferson Lab by impinging a $5.498$-GeV electron beam on a liquid-hydrogen target, and reconstructing the scattered electron and the pion pair with the CLAS detector. 
One-dimensional projections of the $A_{LU}^{\sin\phi_R}$ moments are extracted for the kinematic variables of interest in the valence quark region. 
%The  beam spin asymmetry for hadron-pair production is observed for the first time. %, providing access to quark-gluon correlation.  
The understanding of di-hadron production is essential for the interpretation of observables in single hadron production in semi-inclusive DIS, and pioneering measurements of single spin asymmetries in di-hadron production open a new avenue in studies of QCD dynamics.
\end{abstract}
%PACS: Chromodynamics, quantum; Electromagnetic interactions, Electron-induced nuclear reactions;
\maketitle

The correlations between quarks and gluons occurring inside the nucleon play an essential role in QCD dynamics. Asymmetries from semi-inclusive deep inelastic scattering (SIDIS), where a highly virtual photon interacts with a hadronic target and at least one hadron is detected in the final state, have appeared to be effective tools to access quark distributions and fragmentation information. Studies of hadron pairs in SIDIS open  qualitatively new possibilities to study QCD dynamics, providing access to correlations not accessible with single hadron SIDIS.
%The invariant mass distributions of di-hadrons from different SIDIS and $e^+e^-$ experiments indicate, that a very significant fraction of inclusive pions are coming from correlated di-hadrons. The observables for pions from decays of vector mesons have peculiar spin and momentum dependences and may require different radiative corrections, modeling, and interpretation of observables sensitive to transverse momentum of quarks~\cite{Avakian:2019uzf}.
The interpretation of di-hadron production in SIDIS, as well as interpretation of single-hadron production, are intimately related to contributions to those samples from correlated  di-hadrons in general, and vector mesons, in particular.

At the energy of fixed-target facilities, 
contributions of order  ${\cal O}(M/Q)$, with $M$ the target mass and $Q^2$ the photon virtuality,  become sizeable and, therefore, relevant.
Such contributions are labeled {\it twist-3} effects, and can encode quark-gluon correlations. %At SIDIS energy,  both quark-quark and quark-gluon correlations are accessible. 

In the collinear framework ---in which we shall work in this letter--- %\textit{i.e.} when there is no sensitivity to the transverse dynamics of partons,
six Parton Distribution Functions (PDFs) describe the nucleon up to the twist-3 level. The three leading-twist functions are the unpolarized $f_1(x)$, the helicity  $g_1(x)$ and the tranversity $h_1(x)$ distributions, where $x$ is the Bjorken scaling variable.
The twist-3 PDFs are $e(x)$, $h_L(x)$ and $g_T (x)$, which describe quark-quark and quark-gluon correlations, and as such have no probabilistic interpretation. 

Higher-twist PDFs offer fascinating 
doorways to study the nucleon beyond its valence structure. 
An essential role is played by the {\it chiral-odd} PDF $e(x)$, that is related to the nucleon {\it scalar charge} ---poorly determined through phenomenology--- and {\it a fortiori} to the pion-nucleon sigma term~\cite{Jaffe:1991ra}. 
In addition, it encodes information on the quark mass, as well as genuine quark-gluon correlations~\cite{Efremov:2002qh}. The latter has also been related to an average transverse force acting on a transversely polarized quark in an unpolarized target after the interaction with the virtual photon \cite{Burkardt:2008ps}. It has recently been suggested that this effect could be related to the CP-violating sigma terms~\cite{Seng:2018wwp}. Consequently, the precise measurement of the PDF $e(x)$  could  play an important role in searches for Beyond-the-Standard-Model fundamental scalar interactions~\cite{Bhattacharya:2011qm}, in the same way the tensor charge does~\cite{Courtoy:2015haa}. 

Measurements of twist-3 observables are available from HERMES~\cite{Airapetian:2006rx}, CLAS~\cite{Gohn:2014zbz,Avakian:2003pk,Aghasyan:2011ha} and COMPASS~\cite{Moretti:2019lkw} for single-pion SIDIS, for which the PDF $e(x)$ is accessible only when the transverse momentum of the final hadron is not integrated out.  A first extraction of the --transverse momentum dependent-- $e(x)$ PDF has been pursued from the CLAS data~\cite{Efremov:2002ut}. 
Two-hadron SIDIS measurements provide a method to access twist-3 observables in the collinear framework. 

In this letter, we present the first measurement of the beam spin asymmetry (BSA) in the SIDIS electroproduction of two charged pions using the CLAS data. 
The BSA is defined as the ratio between the difference and the sum of the cross sections corresponding to the two beam-helicity states and can be expressed in terms of harmonics in the azimuthal angle $\phi_R$, defined below. % between the leptonic plane and the plane of transverse component of the relative momentum of the pion pair. 
The $\sin\phi_R$ moment of the BSA can be written as
\begin{equation}
  A_{LU}^{\sin\phi_R} = \sqrt{2\epsilon(1-\epsilon)}\,\frac{F_{LU}^{\sin\phi_R}}{F_{UU,T} + \epsilon F_{UU,L} }\quad,
  \label{ALU}
\end{equation}
with $\epsilon$ the ratio of longitudinal to transverse photon flux, and the structure functions (SFs)~\cite{Bacchetta:2003vn}
% ------- unpolarized SF --------
%
\begin{align} 
\label{F_LUsinphi} 
 F_{LU}^{\sin\phi_R} &=-\sum_q e_q^2\, x\frac{|\bm R| \sin \theta}{Q}\, \nonumber\\
\biggl[
  &  \frac{M}{m_{\pi^+\pi^-}}\,x\, e^q(x)\, H_1^{\open\, q}\bigl(z,\cos \theta, m_{\pi^+\pi^-}\bigr) \nonumber \\ 
  &  +\frac{1}{z}\,f_1^q(x)\,\widetilde{G}^{\open\, q}\bigl(z,\cos \theta, m_{\pi^+\pi^-}\bigr)\biggr]\,,\\
\label{F_UUT}
\hspace{-3mm}
F_{UU ,T} & = \sum_q e_q^2\,x f_1^q(x)\, D_1^q\bigl(z,\cos \theta, m_{\pi^+\pi^-}\bigr), \phantom{\biggl[ \biggr]}
\\
\hspace{-3mm}
F_{UU ,L} &  = 0 ,\phantom{\biggl[ \biggr]}
\end{align} 
contain a sum over the quark flavors $q$ with charge $e_q$, with $M$ the target mass and $m_{\pi^+\pi^-}$ the invariant mass of the pion pair.
The angles are calculated in the center-of-mass frame of the virtual photon-proton system. 
We define the sum of the two pion momenta ${\bf P}_h={\bf P}_{\pi^+}+{\bf P}_{\pi^-}$ as well as their half-difference ${\bf R}=({\bf P}_{\pi^+}-{\bf P}_{\pi^-})/2$.
The azimuthal angle $\phi_{R}$ is defined through the plane formed by the spatial component of ${\bf R}$ orthogonal to ${\bf P}_h$
\begin{eqnarray}
{\bm R}_T&=& \bm{R}-(\bm{R}\cdot {\hat{ \bm P}}_h){\hat {\bm P}}_h\quad,
\end{eqnarray}
and the virtual photon direction, {\it i.e.}
%This leads to the definition of the azimuthal angle
%
\begin{eqnarray}
\phi_{R}&=&  \frac{({\bm q}\times { \bm k})\cdot { \bm R}_T}{|({ \bm q}\times { \bm k})\cdot { \bm R}_T|}\, \arccos ({\hat n}_{\mbox{\tiny lept}}\cdot {\hat n}_{\mbox{\tiny had}})\quad.
\end{eqnarray}
The azimuthal angle $\phi_h$ is defined through the plane formed by ${\bf P}_h$ and the direction of the virtual photon and will be important for the study of acceptance effects.
Both angles, $\phi_h$ and $\phi_R$,
 correspond to the definitions  given in Ref.~\cite{Gliske:2014wba}. 
 
 Each SF has two subscripts indicating the polarization of the beam and target, respectively, and can be written in terms of simple products of a PDF and a di-hadron fragmentation function (FF).
 The SFs depend on the Bjorken variable $x$ (through the PDF) and on the fraction of the virtual photon energy carried by the two hadrons $z=z_{\pi^+}+z_{\pi^-}$, the pion pair invariant mass $m_{\pi^+\pi^-}$ and $\cos\theta$ (through the FF).
Here, $\theta$ is the angle between the direction of one of the final state hadrons in the pair center-of-mass frame and the direction of their centre-of-mass in the photon-target rest frame.
The PDFs and FFs depend also on $Q^2$, through the kinematical suppression of the twist-3 contributions, as well as the QCD evolution equations.

The unpolarized cross section is proportional to the product of the unpolarized PDF $f_1(x)$ and the unpolarized di-hadron FF $D_1$.
We neglect here the structure function $F_{UU}^{\cos\phi_R}$, which is  considered in the systematic uncertainty evaluation only. %because it is expected to give small contribution in the extraction of $A_{LU}$ and is considered in the systematic uncertainty evaluation only. 
The two terms appearing in the polarized structure function $F_{LU}^{\sin\phi_R}$ are both the product of a leading-twist function and a twist-3 one, which could equally contribute.
In Eq.~(\ref{F_LUsinphi}), $e(x)$ appears coupled to the {\it chiral-odd} interference fragmentation function (IFF) $H_1^{\open}$~\cite{Bacchetta:2003vn}. %, analog to the Collins FF in single hadron production.  
%The IFF, at leading order,
%couples to the transversity PDF, which has been accessed~\cite{Bacchetta:2011ip,Bacchetta:2012ty,Radici:2015mwa} through di-hadron target Single-Spin Asymmetry~\cite{dh-transversity,Adolph:2012nw}. 

In the di-hadron framework~\cite{Bianconi:1999cd,Bacchetta:2002ux}, the hadron pair is assumed to be mainly produced in a relative $s$- or $p$-wave channel.
Therefore, it is convenient to expand the $\cos\theta$ dependence of the IFF in a partial wave series that can be truncated to the first-order terms.
This allows for example to replace the IFF $H_1^{\open}(z, \cos\theta, m_{\pi^+\pi^-})$ in Eq.~(\ref{F_LUsinphi}) with the leading term of the partial wave expansion $H_{1,sp}^{\open}(z, m_{\pi^+\pi^-})$, which has been extracted \cite{Courtoy:2012ry,Bacchetta:2012ty,Radici:2015mwa} from Belle data \cite{Vossen:2011fk}.

The analysis we present in this letter is based on the data collected in Hall B of Jefferson Laboratory in 2003 by impinging a $5.498$ GeV longitudinally polarized electron beam on a 5-cm long unpolarized liquid-hydrogen target. 
The CEBAF Large Acceptance Spectrometer (CLAS) \cite{clas} was used to detect the scattered electron and two oppositely charged pions of the reaction $ep\rightarrow e \pi^+ \pi^- X$.
The final di-hadron sample is defined through specific DIS cuts. To be in the scaling regime, the virtuality of the exchanged photon is required to be $Q^2 > 1$ GeV$^2$ and, in order to avoid the resonance region, we impose $W > 2$ GeV. The cut $y < 0.85$ is then applied to suppress radiative events.
Pions coming from the current-fragmentation region were selected by applying on each pion the cut $x_F > $0, where the Feynman-$x$ variable is defined as $x_F=2p_{\parallel}/W$, with $p_{\parallel}$ the pion pair four-momentum component parallel to the virtual-photon direction.
Exclusive events are removed through a cut on the missing mass $m_{e \pi^+\pi^-X} >$ 1.05 GeV.
Spurious contaminations from exclusive baryonic resonance production (for example $\Delta^{++}\pi^-\to p \pi^+\pi^-$) have been studied through Monte Carlo simulations and are further suppressed at the few percent level by cutting on the energy fraction of the pions, namely $z_{\pi^+}>0.28$ and $z_{\pi^-}>0.25$.

Experimentally, the BSA is defined as
\begin{equation}\label{eq_asym}
A_{ LU} = 
\frac{1}{P_B} \frac{ (N_{+} - N_{-} )}{(N_{+} + N_{-} )}\quad ,
\end{equation}
where $N_{+(-)}$ are the number of counts corresponding to each beam-helicity state, and $P_{B} = 0.75 \pm 0.02$ is the average beam polarization over the entire data taking period.
The BSA has been computed in one dimensional projections as a function of $x$, $z$, $m_{\pi^+\pi^-}$ and $Q^2$ and integrating over all the other kinematic variables.

From the theoretical point of view, %after integration over the transverse momentum, 
the only surviving azimuthal modulation of the BSA is the $\sin\phi_R$ moment of Eq.~(\ref{ALU}).
However, kinematic correlations due to the limited phase space available in real data and non-uniform detector acceptance might lead to incomplete cancellation of modulations involving transverse momentum dependent functions~\cite{Bacchetta:2002ux}.
A detailed study based on Monte Carlo simulations demonstrated that a reliable extraction of the $A_{LU}^{\sin\phi_R}$ moment can be achieved by binning the data in a $6\times6$ matrix in the two angles $\phi_R$ and $\Delta\phi=\phi_h-\phi_R$, and performing a 2D fit with the function
\begin{eqnarray}\label{eq_fit_sinphi}
A_{ LU} &=& A_{LU}^{\sin\phi_R}\,\sin\phi_R\nonumber\\
&+&A_{LU}^{\sin(\phi_h-\phi_R)}\,\sin(\phi_h-\phi_R)+A_{LU}^{\sin\phi_h}\,\sin\phi_h
\;,
\end{eqnarray}
representing the relevant modulations from the cross section.

\begin{figure}
\includegraphics[width = 8.5cm]{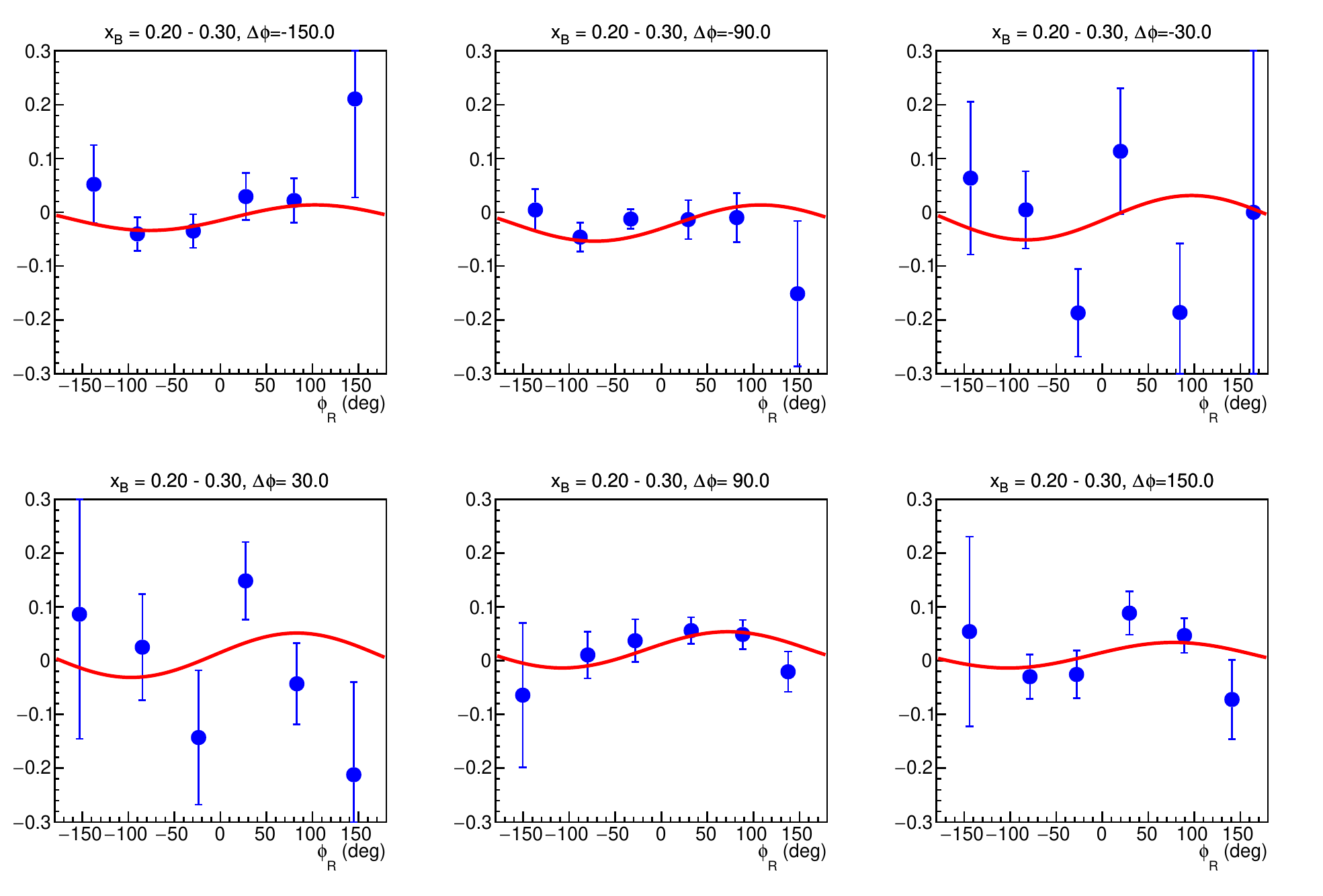}
\caption{(Color online) BSA as a function of $\phi_R$ in the 6 $\Delta\phi$ bins from $-180^o$ to $180^o$ for the bin $x=0.2\div0.3$. The full circles represent the experimental measurement with the vertical bar indicating the total uncertainty, while the curves represent the result of the fit with Eq.~(\ref{eq_fit_sinphi}).}
\vspace{-0.4cm}
\label{fig:ALU_fits}
\end{figure}

An example of the 2D fit in the bin $x=0.2\div0.3$ is shown in Fig. \ref{fig:ALU_fits}. Each panel represents one $\Delta\phi$ bin, the points show the $\phi_R$ dependence of the measured BSA. The curve is the result of the 2D fit.
The fit has been performed considering the total (statistical and systematic) point-to-point uncertainty shown by the error bars.
The systematic uncertainty, amounting to 30\%, is due to the truncation in the partial wave expansion of the di-hadron FFs.
It has been estimated by taking into account the average values of the $\theta$-harmonics of the series from the experimental data and conservatively assuming that the FFs associated to higher harmonics are of the same order of magnitude as the leading one. 
%The error bars include the statistical uncertainty and the point-to-point systematic uncertainty due to the truncation in the partial wave expansion of the di-hadron FFs.
%The latter has been estimated by taking into account the average values of the $\theta$-harmonics of the series from the experimental data and conservatively assuming that the FFs associated to higher harmonics are of the same order of magnitude as the leading one. 
%A 30\% systematic uncertainty has been estimated.
%

\begin{figure}
\includegraphics[width = 8.5cm]{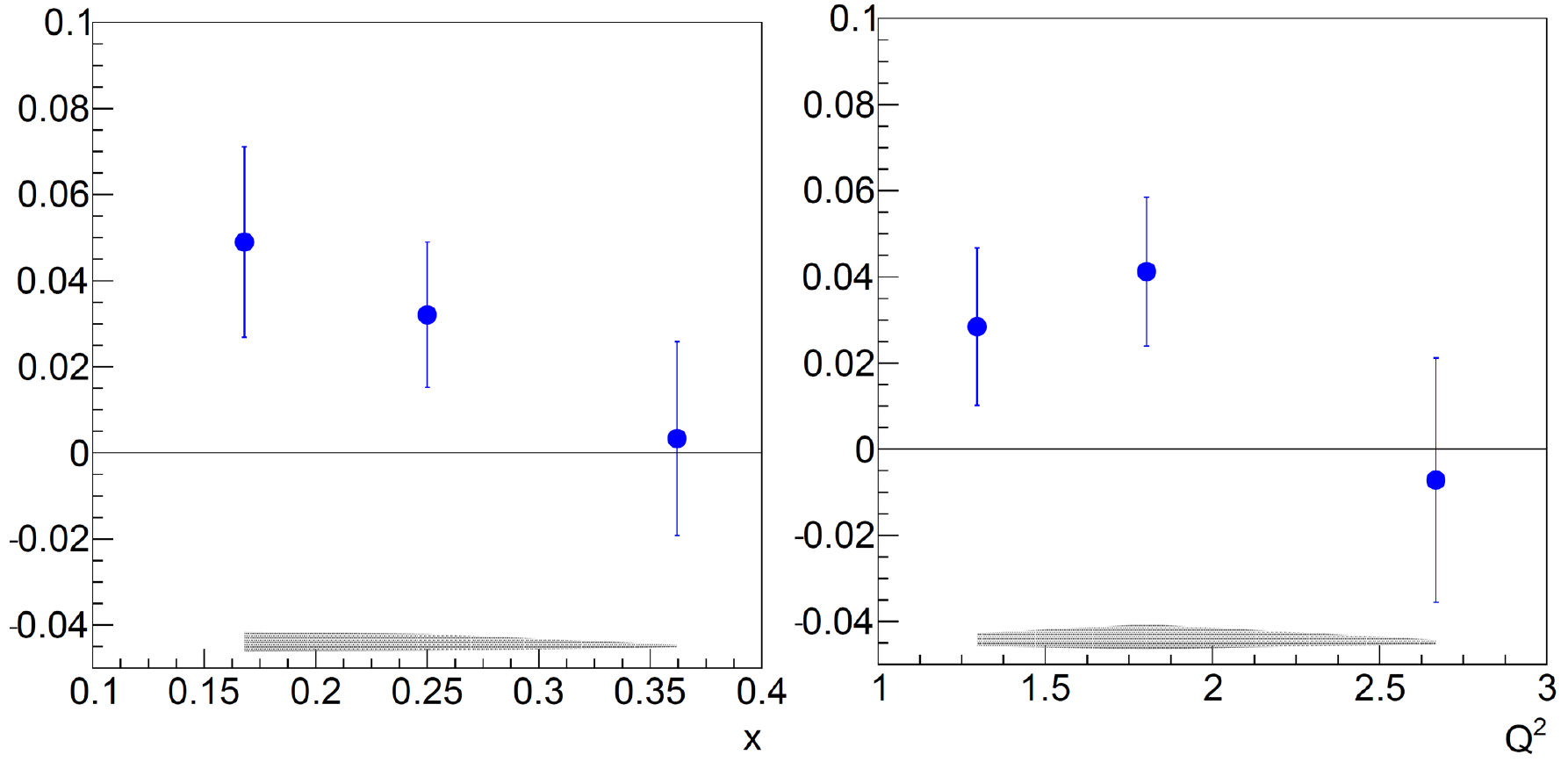}
%\vspace{-0.5cm}
\caption{(Color online) Dependence on $x$ ({\it l.h.s.}) and $Q^2$ ({\it r.h.s.}) of the $\sin\phi_R$ moment $A_{LU}^{\sin\phi_R}$ of the beam spin asymmetry. The error bars and the gray band represents the statistical and systematic uncertainties.}
\vspace{-0.4cm}
\label{fig::bsa_xb}
\label{fig::bsa_Q2}
\end{figure}

\begin{figure}
\includegraphics[width = 8.5cm]{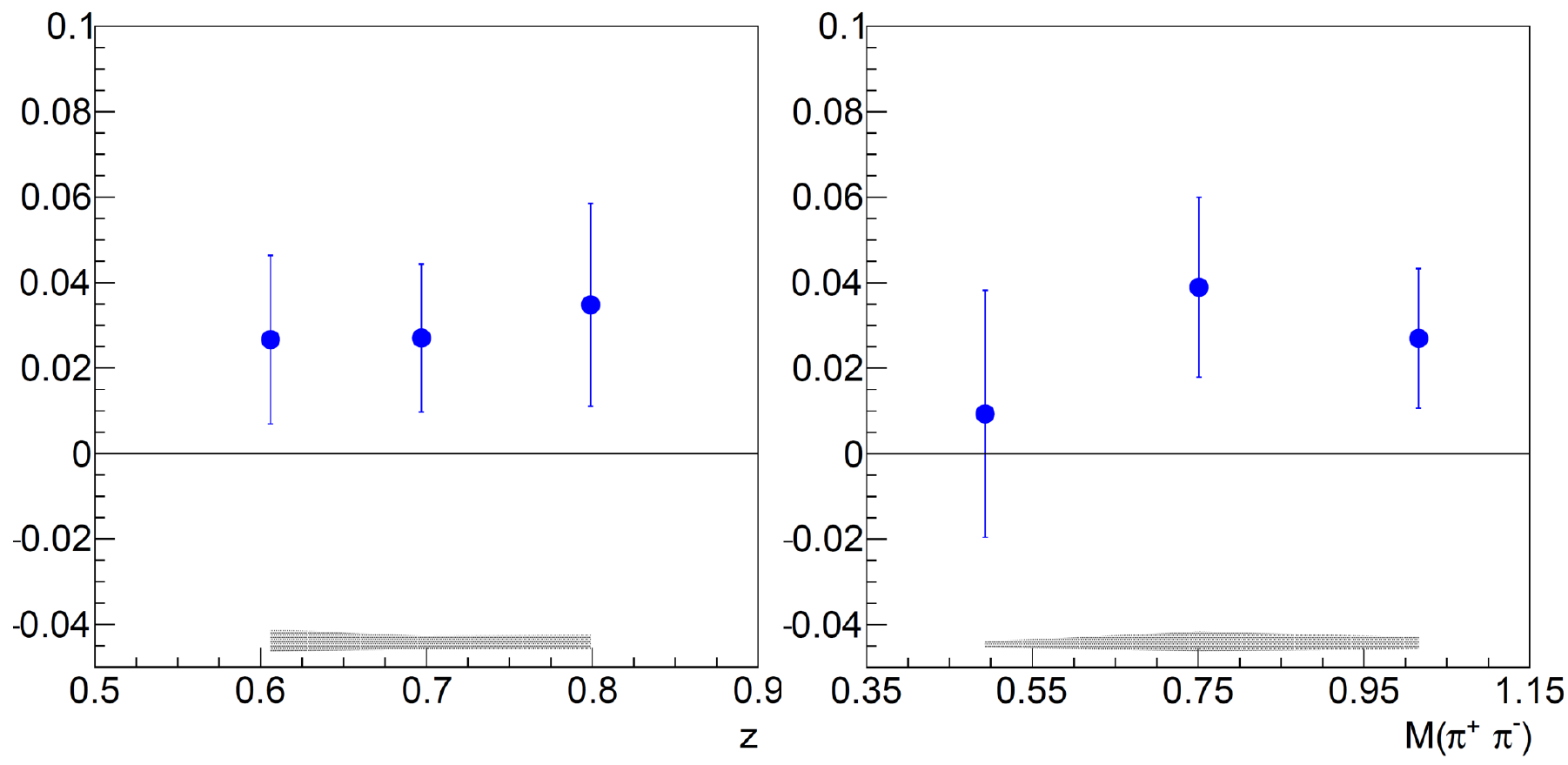}
%\vspace{-0.5cm}
\caption{(Color online) Dependence on $z$ ({\it l.h.s.}) and $m_{\pi\pi}$ ({\it r.h.s.}) of the $\sin\phi_R$ moment $A_{LU}^{\sin\phi_R}$ of the  Asymmetry. The error bars and the gray band represents the statistical and systematic uncertainties.}
\vspace{-0.4cm}
\label{fig::bsa_z_m}
\end{figure}

The $A_{LU}^{\sin\phi_R}$ fitted moments 
are shown in Fig.~\ref{fig::bsa_xb} as a function of $x$ and $Q^2$. 
The projections for $z$ and  $m_{\pi^+\pi^-}$ are given in Fig.~\ref{fig::bsa_z_m}. % (Figs.~\ref{fig::bsa_z} and~\ref{fig::bsa_mh}).
The solid circles correspond to data points. The error bars show the statistical uncertainties from the fits.
At the bottom of each plot, the gray band represents the total systematic uncertainty, which includes:  
a 3\% contribution due to the electron beam polarization;  
a 3\% contribution due to the radiative corrections; 
%a 10\% contribution due to the choice of the fitting function, in particular to the unknown unpolarized $\phi_R$ harmonics; 
the residual contamination from baryon resonance decays, estimated from Monte Carlo studies to be between 2 and 9\% depending on the kinematics.

The kinematic bins in the figures have been chosen so that they have approximately the same statistics, except for $m_{\pi^+\pi^-}$, where the second bin covers the $\rho$ mass region, while the first (third) bin covers the mass region before (after) the $\rho$ mass.
The average $Q^2$ of the data is $1.77$ GeV$^2$.
For the $x$ dependence of the BSA, the invariant mass values range from the threshold to~$\sim 1.7$ GeV and $0.53<z<0.95$. 
%The $z$ projection corresponds to $\langle m_{\pi\pi}\rangle=0.72$ GeV and  $\langle x\rangle=0.27$, and correspondingly for the $m_{\pi\pi}$ projection. 
%, while the projection on the invariant mass bins corresponds to the same averaged values of $x$ and $z$ mentioned above.

% ---------- risultati .........
%
% ---- xb -----
%
%\begin{figure}
%\includegraphics[width = 8cm]{x_ALU-Results_1.pdf}
%%\vspace{-0.5cm}
%\caption{(Color online) Dependence on $x$ of the $\sin\phi_R$ moment $A_{LU}^{\sin\phi_R}$ of the beam spin asymmetry. The error bars and the gray band represents the statistical and systematic uncertainties.}
%\vspace{-0.4cm}
%\label{fig::bsa_xb}
%\end{figure}
%
% ---- xb & Q -----
%
%
%
% ---- Q2 -----
%
%\begin{figure}
%\includegraphics[width = 8cm]{Q2_ALU-Results_2.pdf}
%%\vspace{-0.5cm}
%\caption{(Color online) Dependence on $Q^2$ of the $\sin\phi_R$ moment $A_{LU}^{\sin\phi_R}$ of the beam spin asymmetry. The error bars and the gray band represents the statistical and systematic uncertainties.}
%\vspace{-0.4cm}
%\label{fig::bsa_Q2}
%\end{figure}
%
% ---- Q2 -----
%

\begin{figure}
\includegraphics[width = 8cm]{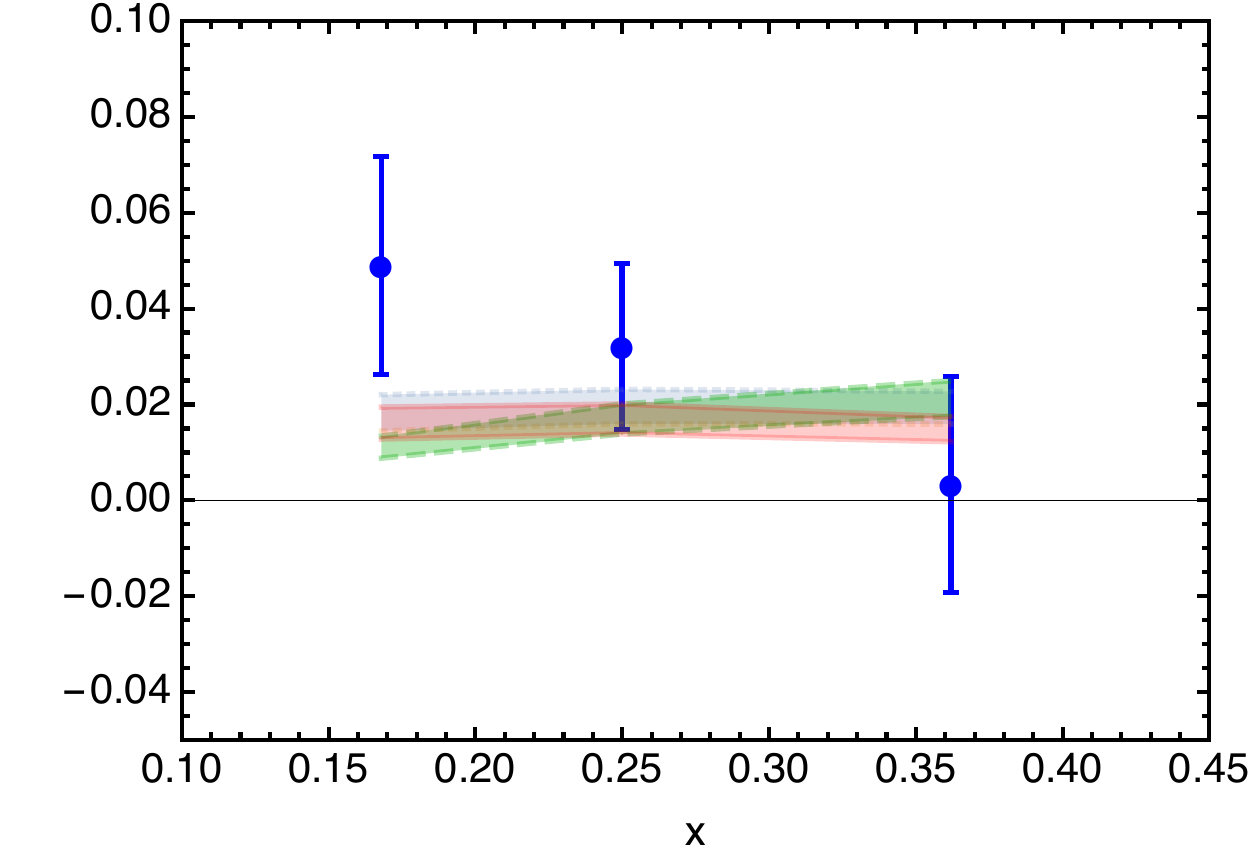}
%\vspace{-0.5cm}
\caption{(Color online) Dependence on $x$ of the $\sin\phi_R$ moment $A_{LU}^{\sin\phi_R}$ of the beam spin asymmetry compared to model calculations: the spectator model~\cite{Jakob:1997wg} in green-dashed, the bag model~\cite{Signal:1996ct} in red-full and the gray-dotted band corresponds to the LFCQM~\cite{Pasquini:2018oyz}, all using the di-hadron FFs~\cite{Courtoy:2012ry}. 
}
\vspace{-0.4cm}
\label{fig::bsa_models}
\end{figure}

%
% ------ previous measurements -------
%
%
In this letter we report the pioneering observation of a twist-3 observable in di-hadron SIDIS; hence, no previous measurements are available for comparison. 
The $x$ projection of the $\sin\phi_R$ moment reflects the behavior of the PDFs. %However, some qualitative features of di-hadron SIDIS can be noted. The IFFs  are dominated by an interference between a $s$-wave contribution from the incoherent part of the pion-pair spectrum and the $p$-wave coming from the $\rho$ channel~\cite{Jaffe:1997hf,Bacchetta:2002ux}. This scheme, first used in Ref.~\cite{Bacchetta:2006un}, has been tested and confirmed in Ref.~\cite{Courtoy:2012ry} as well as by other experiments. %The trend of the ratio $H_1^{\sphericalangle}(z, m_{\pi^+\pi^-})/D_1(z, m_{\pi^+\pi^-})$ as observed in Figs.~\ref{fig::bsa_z}-\ref{fig::bsa_mh} is in agreement with its phenomenological determination~\cite{Courtoy:2012ry,Radici:2015mwa}. 
In Fig.~\ref{fig::bsa_models} this projection of $A_{LU}^{\sin\phi_R}$ is compared to models for the twist-3 PDF $e(x)$ ---neglecting the second term on the {\it r.h.s} of Eq.~(\ref{F_LUsinphi}). The curves corresponding to the different models for $e(x)$ are produced by combining to the extracted di-hadron FFs~\cite{Courtoy:2012ry}.  
%and evolving it to the $Q^2$ of the present measurement with the model predictions for $e(x)$.
The gray-dotted band corresponds to the LFCQM of %including gluon degrees of freedom through higher Fock states
Ref.~\cite{Pasquini:2018oyz} together with MSTW08LO unpolarized PDF~\cite{Martin:2009iq}, the green-dashed band corresponds to the asymmetry with both PDFs evaluated in the spectator model~\cite{Jakob:1997wg}, while the red-full band corresponds to the MIT bag model~\cite{Signal:1996ct} for both $f_1(x)$ and $e(x)$. 
Other calculations, not shown in the figure, have been proposed~\cite{Ohnishi:2003mf,Schweitzer:2003uy,Lorce:2014hxa}.
The models are consistent among themselves; they are in agreement with the experimental data within about 1$\sigma$, although the data seems to indicate a steeper decrease at high $x$ than the models.
%No conclusions about model predictions can be driven due to the size of the uncertainty of our measurement.
%
The behavior of the three points could be examined in a thorough point-by-point extraction of $e(x)$~\cite{us:tocome}, as has been sketched in Ref.~\cite{Courtoy:2014ixa} using preliminary results from the same CLAS data we present in this letter.
\\

In summary, for the first time a beam spin asymmetry for the semi-inclusive electroproduction of charged pion pairs has been measured. 
The one-dimensional projections of the moment $A_{LU}^{\sin\phi_R}$ in $x$, $z$, $m_{\pi^+\pi^-}$ and $Q^2$ have been extracted. 
This measurement constitutes a pioneering study that has accounted for systematic sources of uncertainties that were shown to have little impact on the present results but that would be crucial %to understand in high statistics experiments
%This measurement constitutes an important reference 
for ongoing studies of di-hadron observables using the CLAS12 \cite{CLAS2} detector with proton and deuteron targets~\cite{our_proposal}, at an order of magnitude higher luminosity, which will provide further information on the twist-3 nucleon structure in the mid-$x$ region. 

The invariant mass distributions of di-hadrons from different SIDIS and $e^+e^-$ experiments indicate that a very significant fraction of inclusive pions are coming from correlated di-hadrons. The observables for pions from decays of vector mesons have peculiar spin and momentum dependences and may require different radiative corrections, modeling, and interpretation of observables sensitive to the transverse momentum of quarks~\cite{Avakian:2019uzf}. As such, di-hadron measurements establish the bases for future experimental and phenomenological studies.
\\

%--------------------------------------------------------
\begin{acknowledgements}
We thank the staff of the Accelerator and Physics Divisions for making the experiment possible. Special thanks to A. Bacchetta, B. Pasquini, M. Polyakov, M. Radici, P. Schweitzer and M. Wakamatsu for useful discussions. This work was supported in part by the U.S. Department of Energy (No. DE-FG02-96ER40950) and National Science Foundation, the French Centre National de la Recherche Scientifique and Commissariat  \`a l'Energie Atomique, the French-American Cultural Exchange (FACE), the Italian Istituto Nazionale di Fisica Nucleare, the Chilean Comisi\'on Nacional de Investigaci\'on Cient\'ifica y Tecnol\'ogica (CONICYT), the Mexican Consejo Nacional de Ciencias y Tecnolog\'ia (CONACyT), the National Research Foundation of Korea, and the UK Science and Technology Facilities Council (STFC). A.C. is supported by DGAPA-PAPIIT IA101720 and CONACyT Ciencia de Frontera 2019 No.~51244 (FORDECYT-PRONACES). The Jefferson Science Associates (JSA) operates the Thomas Jefferson National Accelerator Facility for the United States Department of Energy under contract DE-AC05-06OR23177.
\end{acknowledgements}

%
%
% --------- bibliography -----------
%

\clearpage
%\newpage
%\appendix

%\begin{widetext}
\onecolumngrid
\section{Supplemental material}
\label{Sect:suppl}

We provide in this Section Tables that support the main part of the manuscript. In Table~\ref{tab:bins}, the kinematic ranges for the relevant variables are given. The full kinematic ranges are given in Tables~\ref{tab:bins_xB}, \ref{tab:bins_Q}, \ref{tab:bins_z} and \ref{tab:bins_Mh}; the asymmetries, statistical and systematical uncertainties are provided for each 1D projection in Tables~\ref{tab:BSA_x}, \ref{tab:BSA_Q}, \ref{tab:BSA_z} and \ref{tab:BSA_mh}. 

%%%%%%%%%%%%%TABLE KINEMATICS%%%%%%%%%%%%%%%%%%
\begin{table}[h]
\begin{center}
\begin{tabular}{||p{2.5cm}||p{2cm}|p{2cm}|p{2cm}||} \hline \hline
Variable & Bin 1 & Bin 2 & Bin 3   \\ \hline
\hline
$x$ &   0.114$\div$0.200 &  0.200$\div$0.300  &  0.300$\div$0.593  \\ \hline
$Q^2$ (GeV$^2$) &   1.000$\div$1.500 &  1.500$\div$2.200  &  2.200$\div$4.644  \\ \hline
$z$ &   0.530$\div$0.650 &  0.650$\div$0.750  &  0.705$\div$0.948  \\ \hline
$m_{\pi^+\pi^-}$ (GeV) &   0.279$\div$0.650 &  0.650$\div$0.852  &  0.852$\div$1.734  \\ \hline
\hline
\end{tabular}
\end{center}
\caption{ Bins of the 1D projections of the $A_{LU}^{\sin\phi_R}$ measurements for each of the respective projections.}
\label{tab:bins}
\end{table}
%%%%%%%%%%%%%TABLE values xB%%%%%%%%%%%%%%%%%%
\begin{table}[h]
\begin{center}
  \begin{tabular}{||c||p{1.15cm}|p{1.15cm}|p{1.15cm}||p{1.15cm}|p{1.15cm}|p{1.15cm}||p{1.15cm}|p{1.15cm}|p{1.15cm}||p{1.15cm}|p{1.15cm}|p{1.15cm}||} \hline \hline
     &  \multicolumn{3}{|c||}{$x$} &  \multicolumn{3}{|c||}{$Q^2$ (GeV$^2$)} & \multicolumn{3}{|c||}{$z$} &  \multicolumn{3}{|c||}{$m_{\pi^+\pi^-}$ (GeV)} \\ \hline 
    bin & min & max & ave & min & max & ave & min & max & ave & min & max & ave \\ \hline \hline
    1 & \footnotesize 0.114 & \footnotesize 0.200 & \footnotesize 0.168 & \footnotesize 1.000 & \footnotesize 1.752 & \footnotesize 1.279 & \footnotesize 0.531 & \footnotesize 0.948 & \footnotesize 0.692 & \footnotesize 0.279 & \footnotesize 1.734 & \footnotesize 0.826  \\ \hline 
      2 & \footnotesize 0.200 & \footnotesize 0.300 & \footnotesize 0.250 & \footnotesize 1.084 & \footnotesize 2.628 & \footnotesize 1.673 & \footnotesize 0.530 & \footnotesize 0.933 & \footnotesize 0.691 & \footnotesize 0.279 & \footnotesize 1.627 & \footnotesize 0.718 \\ \hline 
      3 & \footnotesize 0.300 & \footnotesize 0.593 & \footnotesize 0.362 & \footnotesize 1.356 & \footnotesize 4.644 & \footnotesize 2.353 & \footnotesize 0.531 & \footnotesize 0.913 & \footnotesize 0.680 & \footnotesize 0.279 & \footnotesize 1.384 & \footnotesize 0.626 \\ \hline \hline
\end{tabular}
\end{center}
\caption{ Kinematic ranges of the three $x$ bins.}
\label{tab:bins_xB}
\end{table}

%%%%%%%%%%%%BSA x%%%%%%%%
\begin{table}[h]
\begin{center}
\begin{tabular}{||p{1cm}||p{1.5cm}|p{1.75cm}|p{1.5cm}|p{1.5cm}||} \hline \hline
Bin & $<x>$ & $A_{LU}^{\sin\phi_R}(x)$ & Stat & Syst   \\ \hline
\hline
1 &   0.168 &  0.0490  &  0.0221  &  0.0054  \\ \hline
2 &   0.250 &  0.0321  &  0.0169  &  0.0038  \\ \hline
3 &   0.362 &  0.0033  &  0.0225  &  0.0004  \\ \hline
\hline
\end{tabular}
\end{center}
\caption{ Asymmetries for the $x$ projection.}
\label{tab:BSA_x}
\end{table}
%%%%%%%%%%%%%TABLE values Q2%%%%%%%%%%%%%%%%%%
\begin{table}[h]
\begin{center}
  \begin{tabular}{||c||p{1.15cm}|p{1.15cm}|p{1.15cm}||p{1.15cm}|p{1.15cm}|p{1.15cm}||p{1.15cm}|p{1.15cm}|p{1.15cm}||p{1.15cm}|p{1.15cm}|p{1.15cm}||} \hline \hline
     &  \multicolumn{3}{|c||}{$x$} &  \multicolumn{3}{|c||}{$Q^2$ (GeV$^2$)} & \multicolumn{3}{|c||}{$z$} &  \multicolumn{3}{|c||}{$m_{\pi^+\pi^-}$ (GeV)} \\ \hline 
    bin & min & max & ave & min & max & ave & min & max & ave & min & max & ave \\ \hline \hline
    1 & \footnotesize 0.114 & \footnotesize 0.324 & \footnotesize 0.197 & \footnotesize 1.000 & \footnotesize 1.500 & \footnotesize 1.296 & \footnotesize 0.531 & \footnotesize 0.948 & \footnotesize 0.693 & \footnotesize 0.279 & \footnotesize 1.734 & \footnotesize 0.748  \\ \hline 
      2 & \footnotesize 0.171 & \footnotesize 0.412 & \footnotesize 0.274 & \footnotesize 1.500 & \footnotesize 2.200 & \footnotesize 1.803 & \footnotesize 0.530 & \footnotesize 0.938 & \footnotesize 0.688 & \footnotesize 0.279 & \footnotesize 1.627 & \footnotesize 0.704 \\ \hline 
      3 & \footnotesize 0.251 & \footnotesize 0.593 & \footnotesize 0.374 & \footnotesize 2.200 & \footnotesize 4.644 & \footnotesize 2.668 & \footnotesize 0.531 & \footnotesize 0.924 & \footnotesize 0.678 & \footnotesize 0.279 & \footnotesize 1.492 & \footnotesize 0.673\\ \hline \hline
\end{tabular}
\end{center}
\caption{ Kinematic ranges of the three $Q^2$ bins.}
\label{tab:bins_Q}
\end{table}
%%%%%%%%%%%BSA Q2%%%%%%%%
\begin{table}[h]
\begin{center}
\begin{tabular}{||p{1cm}||p{2.5cm}|p{1.7cm}|p{1.5cm}|p{1.5cm}||} \hline \hline
Bin & $<Q^2>$ (GeV$^2$) & $A_{LU}^{\sin\phi_R}(Q^2)$ & Stat & Syst   \\ \hline
\hline
1 &  1.296 &  0.0284  &  0.0183  &  0.0032  \\ \hline
2 &  1.803 &  0.0412  &  0.0173  &  0.0051  \\ \hline
3 &  2.668 & -0.0072  &  0.0284  &  0.0009  \\ \hline
\hline
\end{tabular}
\end{center}
\caption{ Asymmetries for the $Q^2$ projection.}
\label{tab:BSA_Q}
\end{table}
%%%%%%%%%%%%%TABLE values z%%%%%%%%%%%%%%%%%%
\begin{table}[h]
\begin{center}
  \begin{tabular}{||c||p{1.15cm}|p{1.15cm}|p{1.15cm}||p{1.15cm}|p{1.15cm}|p{1.15cm}||p{1.15cm}|p{1.15cm}|p{1.15cm}||p{1.15cm}|p{1.15cm}|p{1.15cm}||} \hline \hline
     &  \multicolumn{3}{|c||}{$x$} &  \multicolumn{3}{|c||}{$Q^2$ (GeV$^2$)} & \multicolumn{3}{|c||}{$z$} &  \multicolumn{3}{|c||}{$m_{\pi^+\pi^-}$ (GeV)} \\ \hline 
    bin & min & max & ave & min & max & ave & min & max & ave & min & max & ave \\ \hline \hline
    1 & \footnotesize 0.114 & \footnotesize 0.593 & \footnotesize 0.271 & \footnotesize 1.000 & \footnotesize 4.588 & \footnotesize 1.834 & \footnotesize 0.530 & \footnotesize 0.650 & \footnotesize 0.606 & \footnotesize 0.279 & \footnotesize 1.383 & \footnotesize 0.660  \\ \hline 
      2 & \footnotesize 0.114 & \footnotesize 0.592 & \footnotesize 0.272 & \footnotesize 1.000 & \footnotesize 4.644 & \footnotesize 1.820 & \footnotesize 0.650 & \footnotesize 0.750 & \footnotesize 0.697 & \footnotesize 0.279 & \footnotesize 1.653 & \footnotesize 0.727 \\ \hline 
      3 & \footnotesize 0.114 & \footnotesize 0.551 & \footnotesize 0.255 & \footnotesize 1.000 & \footnotesize 4.176 & \footnotesize 1.723 & \footnotesize 0.750 & \footnotesize 0.948 & \footnotesize 0.799 & \footnotesize 0.279 & \footnotesize 1.734 & \footnotesize 0.770\\ \hline \hline
\end{tabular}
\end{center}
\caption{ Kinematic ranges of the three $z$ bins.}
\label{tab:bins_z}
\end{table}
%%%%%%%%%%%BSA z%%%%%%%%
\begin{table}[h]
\begin{center}
\begin{tabular}{||p{1cm}||p{1.75cm}|p{1.5cm}|p{1.5cm}|p{1.5cm}||} \hline \hline
Bin & $<z>$ & $A_{LU}^{\sin\phi_R}(z)$ & Stat & Syst   \\ \hline
\hline
1 &  0.606 &  0.0267  &  0.0197  &  0.0037  \\ \hline
2 &  0.697 &  0.0270  &  0.0173  &  0.0031  \\ \hline
3 &  0.799 &  0.0348  &  0.0237  &  0.0039  \\ \hline
\hline
\end{tabular}
\end{center}
\caption{ Asymmetries for the $z$ projection.}
\label{tab:BSA_z}
\end{table}
%%%%%%%%%%%%%TABLE values mpipi%%%%%%%%%%%%%%%%%%
\begin{table}[h]
\begin{center}
  \begin{tabular}{||c||p{1.15cm}|p{1.15cm}|p{1.15cm}||p{1.15cm}|p{1.15cm}|p{1.15cm}||p{1.15cm}|p{1.15cm}|p{1.15cm}||p{1.15cm}|p{1.15cm}|p{1.15cm}||} \hline \hline
     &  \multicolumn{3}{|c||}{$x$} &  \multicolumn{3}{|c||}{$Q^2$ (GeV$^2$)} & \multicolumn{3}{|c||}{$z$} &  \multicolumn{3}{|c||}{$m_{\pi^+\pi^-}$ (GeV)} \\ \hline 
    bin & min & max & ave & min & max & ave & min & max & ave & min & max & ave \\ \hline \hline
    1 & \footnotesize 0.114 & \footnotesize 0.593 & \footnotesize 0.291 & \footnotesize 1.000 & \footnotesize 4.644 & \footnotesize 1.865 & \footnotesize 0.530 & \footnotesize 0.933 & \footnotesize 0.676 & \footnotesize 0.279 & \footnotesize 0.650 & \footnotesize 0.493\\ \hline 
    2 & \footnotesize 0.114 & \footnotesize 0.590 & \footnotesize 0.267 & \footnotesize 1.000 & \footnotesize 4.588 & \footnotesize 1.805 & \footnotesize 0.531 & \footnotesize 0.940 & \footnotesize 0.687 & \footnotesize 0.650 & \footnotesize 0.852 & \footnotesize 0.751\\ \hline 
    3 & \footnotesize 0.114 & \footnotesize 0.550 & \footnotesize 0.231 & \footnotesize 1.000 & \footnotesize 4.344 & \footnotesize 1.705 & \footnotesize 0.531 & \footnotesize 0.948 & \footnotesize 0.706 & \footnotesize 0.850 & \footnotesize 1.734 & \footnotesize 1.016 \\ \hline \hline
\end{tabular}
\end{center}
\caption{ Kinematic ranges of the three $m_{\pi^+\pi^-}$ bins.}
\label{tab:bins_Mh}
\end{table}
%%%%%%%%%%%BSA z%%%%%%%%
\begin{table}[h]
\begin{center}
\begin{tabular}{||p{1cm}||p{2.75cm}|p{2.3cm}|p{1.5cm}|p{1.5cm}||} \hline \hline
Bin & $<m_{\pi^+\pi^-}>$ (GeV)& $A_{LU}^{\sin\phi_R}(m_{\pi^+\pi^-})$ & Stat & Syst   \\ \hline
\hline
1 &  0.493  & 0.0093  &  0.0289  &  0.0012  \\ \hline
2 &  0.751  & 0.0389  &  0.0211  &  0.0045  \\ \hline
3 &  1.016  & 0.0270  &  0.0163  &  0.0030  \\ \hline
\hline
\end{tabular}
\end{center}
\caption{ Asymmetries for the $m_{\pi^+\pi^-}$ projection.}
\label{tab:BSA_mh}
\end{table}
%%%%%%%%%%%%%%%%%%%%%%%%%%%%%%%
%
%
%
%

% ---- z -----
%
%\begin{figure}[h]
%\includegraphics[width = 8cm]{z_ALU-Results_4.pdf}
%%\vspace{-0.5cm}
%\caption{(Color online) Dependence on $z$ of the $\sin\phi_R$ moment $A_{LU}^{\sin\phi_R}$ of the beam spin asymmetry. The error bars and the gray band represents the statistical and systematic uncertainties.}
%\vspace{-0.4cm}
%\label{fig::bsa_z}
%\end{figure}

% ---- mh -----
%
%\begin{figure}[h]
%\includegraphics[width = 8cm]{mh_ALU-Results_5.pdf}
%%\vspace{-0.5cm}
%\caption{(Color online) Dependence on $m_{\pi^+\pi^-}$ of the $\sin\phi_R$ moment $A_{LU}^{\sin\phi_R}$ of the beam spin asymmetry. The error bars and the gray band represents the statistical and systematic uncertainties. }
%\vspace{-0.4cm}
%\label{fig::bsa_mh}
%\end{figure}
%
%
%
%\end{widetext}

\end{document}